\documentclass[twocolumn,showpacs,showkeys,preprintnumbers,amsmath,amssymb]{revtex4}

 \usepackage{graphicx}
\usepackage{amssymb}
\usepackage{float}
\usepackage{dcolumn}% Align table columns on decimal point
\usepackage{bm}% bold math

\floatstyle{ruled}
\newfloat{algorithm}{htbp}{loa}%[section] %[chapter]
\floatname{algorithm}{Algorithm}  % OPTION OF "H" meaning HERE -> check it out

\newtheorem{defn2}{Definition}% [section]

\newlength\figwidth
\begin{document}

\title{Huffman Coding as a Non-linear Dynamical System}

\author{Nithin Nagaraj}%
 \email{nithin_nagaraj@yahoo.com}\homepage{http://nithin.nagaraj.googlepages.com}
\affiliation{Department of Electronics and Communication Engg.,
Amrita School of Engg., Amrita Vishwa Vidyapeetham, Amritapuri
Campus, Kerala 690 525, India.}

\date{\today}

\begin{abstract}
In this paper, source coding or data compression is viewed as a
measurement problem. Given a measurement device with fewer states
than the observable of a stochastic source, how can one capture the
essential information? We propose modeling stochastic sources as
piecewise linear discrete chaotic dynamical systems known as
Generalized Lur\"{o}th Series (GLS) which dates back to Georg
Cantor's work in 1869. The Lyapunov exponent of GLS is equal to the
Shannon's entropy of the source (up to a constant of
proportionality). By successively approximating the source with GLS
having fewer states (with the closest Lyapunov exponent), we derive
a binary coding algorithm which exhibits minimum redundancy (the
least average codeword length with integer codeword lengths). This
turns out to be a re-discovery of Huffman coding, the popular
lossless compression algorithm used in the JPEG international
standard for still image compression.
\end{abstract}

\pacs{89.70.Cf, 02.50.-r, 87.19.lo}

\keywords{ Source coding, Data compression, Generalized Lur\"{o}th
Series, Chaotic map, Huffman coding}

\maketitle

\section{Source Coding seen as a Measurement Problem}
\label{Section:Measurement Problem} 
\begin{figure}[!h]
\centering
\includegraphics[scale=0.4]{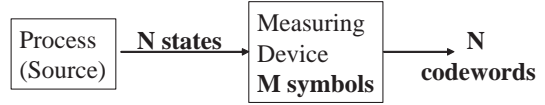}
\caption[Source coding as a measurement problem]{Source coding as a
measurement problem. Typically, $M \ll N$. If $M=2$, we are seeking
binary codes.} \label{figure:measurement}
\end{figure}

A practical problem an
experimental physicist would face is the following -- a process (eg.
a particle moving in space) has an observed variable (say position
of the particle) which potentially takes $N$ distinct values, but
the measuring device is capable of recording only $M$ values  and
$M<N$. In such a scenario (Figure~\ref{figure:measurement}), how can
we make use of these $M$ states of the measuring device to capture
the essential information of the source? It may be the case that $N$
takes values from an infinite set, but the measuring device is
capable of recording only a finite number of states. However, it
shall be assumed that $N$ is finite but allowed for the possibility
that $N \gg M$ (for e.g., it is possible that $N=10^6$ and $M=2$).

Our aim is to capture the essential information of the source (the
process is treated as a source and the observations as messages from
the source) in a lossless fashion. This problem actually goes all
the way back to Shannon~\cite{Shannon1948} who gave a mathematical
definition for the information content of a source. He defined it as
`Entropy', a term borrowed from statistical thermodynamics.
Furthermore, his now famous noiseless source coding theorem states
that it is possible to encode the information of a memoryless source
(assuming that the observables are independent and identically
distributed (i.i.d)) using (at least) $H(X)$ bits per symbol, where
$H(X)$ stands for the Shannon's entropy of the source $X$. Stated in
other words, the average codeword length $\sum_{i}^{ } l_i p_i \leq
H(X)$ where $l_i$ is the length of the $i$-th codeword and $p_i$ the
corresponding probability of occurrence of the $i$-th alphabet of
the source.

Shannon's entropy $H(X)$ defines the ultimate limit for lossless
data compression. Data compression is a very important and exciting
research topic in Information theory since it not only provides a
practical way to store bulky data, but it can also be used
effectively to measure entropy, estimate complexity of sequences and
provide a way to generate pseudo-random numbers~\cite{MacKay} (which
are necessary for Monte-Carlo simulations and Cryptographic
protocols).

Several researchers have investigated the relationship between
chaotic dynamical systems and data compression (more generally
between chaos and information theory). Jim\'{e}nez-Monta\~{n}o,
Ebeling, and others~\cite{nsrps1} have proposed coding schemes by a
symbolic substitution method. This was shown to be an optimal data
compression algorithm by Grassberger~\cite{Grassberger} and also to
accurately estimate Shannon's entropy~\cite{Grassberger} and
Lyapunov exponents of dynamical systems~\cite{nsrps2}. Arithmetic
coding, a popular data compression algorithm used in JPEG2000 was
recently shown to be a specific mode of a piecewise linear chaotic
dynamical system~\cite{NithinAC}. In another
work~\cite{NithinKraft}, we have used symbolic dynamics on chaotic
dynamical systems to prove the famous Kraft-McMillan inequality and
its converse for prefix-free codes, a fundamental inequality in
source coding, which also has a Quantum analogue.

In this paper, we take a nonlinear dynamical systems approach to the
aforementioned measurement problem. We are interested in modeling
the source by a nonlinear dynamical system. By a suitable model, we
hope to capture the information content of the source. This paper is
organized as follows. In Section II, stochastic sources are modeled
using piecewise linear chaotic dynamical systems which exhibits some
important and interesting properties. In Section III, we propose a
new algorithm for source coding and prove that it achieves the least
average codeword length and turns out to be a re-discovery of
Huffman coding~\cite{Huffman1952} -- the popular lossless
compression algorithm used in the JPEG international
standard~\cite{JPEG} for still image compression. We make some
observations about our approach in Section IV and conclude in
Section V.

\section{Source Modeling using Piecewise Linear Chaotic Dynamical Systems}
We shall consider stationary sources. These are defined as sources
whose statistics remain constant with respect to
time~\cite{Papoulis2002}. These include independent and identically
distributed (i.i.d) sources and Ergodic (Markov) sources. These
sources are very important in modeling various
physical/chemical/biological phenomena and in engineering
applications~\cite{Kohda2002}.

On the other hand, non-stationary sources are those whose statistics
change with time. We shall not deal with them here. However, most
coding methods are applicable to these sources with some suitable
modifications.

\subsection{Embedding an i.i.d Source using Generalized Lur\"{o}th
Series} \label{Subsection:embedding}
 Consider an i.i.d source $X$ (treated as a random variable)
which takes values from a set of $N$ values $\mathcal{A} = \{a_1,
a_2, \ldots, a_N  \}$ with probabilities $\{ p_1, p_2, \ldots, p_N
\}$ respectively with the condition $\sum_{1}^{N} p_i = 1$.

An i.i.d source can be simply modeled as a (memoryless) Markov
source (or Markov process~\cite{Papoulis2002}) with the transition
probability from state $i$ to $j$ as being independent of state $i$
(and all previous states)~\footnote{In other words, $P(X_{n+1} = j |
X_{n} = i, X_{n-1} = h,\ldots, X_{1}=a) = P(X_{n+1} = j) = p_j$.}.
We can then embed the Markov source into a dynamical system as
follows: to each Markov state (i.e. to each symbol in the alphabet),
associate an interval on the real line segment $[0,1)$ such that its
length is equal to the probability. Any two such intervals have
pairwise disjoint interiors and the union of all the intervals cover
$[0,1)$. Such a collection of intervals is known as a partition. We
define a deterministic map $T$ on the partitions such that they form
a Markov partition (they satisfy the property that the image of each
interval under $T$ covers an integer number of
partitions~\cite{Afraimovich2003}).

\begin{figure}[!h]
\centering
\includegraphics[scale=0.42]{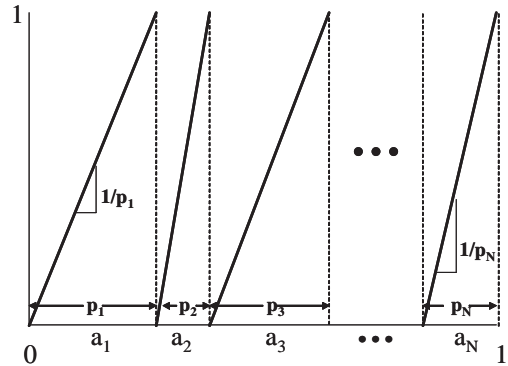}\\
(a) Generalized Lur\"{o}th Series (GLS) $T: [0,1) \mapsto [0,1)$
\includegraphics[scale=0.42]{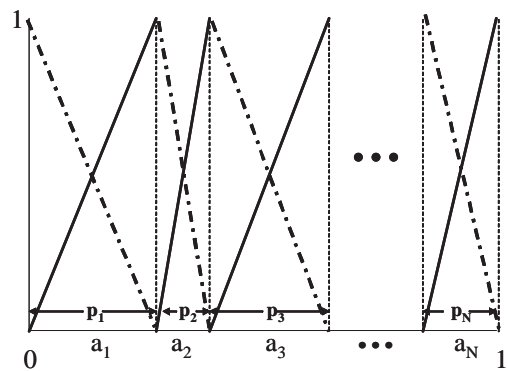}\\
(b) $2^N$ modes of GLS. \caption[Embedding an i.i.d source into a
Generalized Lur\"{o}th Series]{Embedding an i.i.d source into a
Generalized Lur\"{o}th Series (GLS).} \label{figure:GLSandModes}
\end{figure}

The simplest way to define the map $T$ such that the intervals form
a Markov partition is to make it linear and surjective. This is
depicted in Figure~\ref{figure:GLSandModes}(a). Such a map is known
as Generalized Lur\"{o}th Series (GLS). There are other ways to
define the map $T$ (for eg., see~\cite{Kohda2002}) but for our
purposes GLS will suffice. Lur\"{o}th's paper in 1883 (see reference
in Dajani et. al.~\cite{Dajani2002}) deals with number theoretical
properties of Lur\"{o}th series (a specific case of GLS). However,
Georg Cantor had discovered GLS earlier in 1869~\cite{Waterman1975,
Cantor1869}.

\subsection{Some Important Properties of GLS}
\label{subsec:imppropGLS} A list of important properties of GLS is
given below:

\begin{enumerate}
\item GLS preserves the Lebesgue (probability) measure.
\item Every (infinite) sequence of symbols from the alphabet corresponds to an unique initial
condition.
\item The symbolic sequence of every initial condition is i.i.d.
\item GLS is Chaotic (positive Lyapunov exponent, positive Topological
entropy).
\item GLS has maximum topological entropy ($=ln(N)$) for a specified number of alphabets ($N$). Thus, all possible
arrangements of the alphabets can occur as symbolic sequences.
\item GLS is isomorphic to the Shift map and hence Ergodic (Bernoulli).
\item Modes of GLS: As it can be seen from
Figure~\ref{figure:GLSandModes}(b), the slope of the line that maps
each interval to $[0,1)$ can be {\it chosen} to be either positive
or negative. These choices result in a total of $2^N$ {\it modes} of
GLS (up to a permutation of the intervals along with their
associated alphabets for each mode, these are $N!$ in number).
\end{enumerate}

\begin{figure}[!h]
\centering
\includegraphics[scale=0.25]{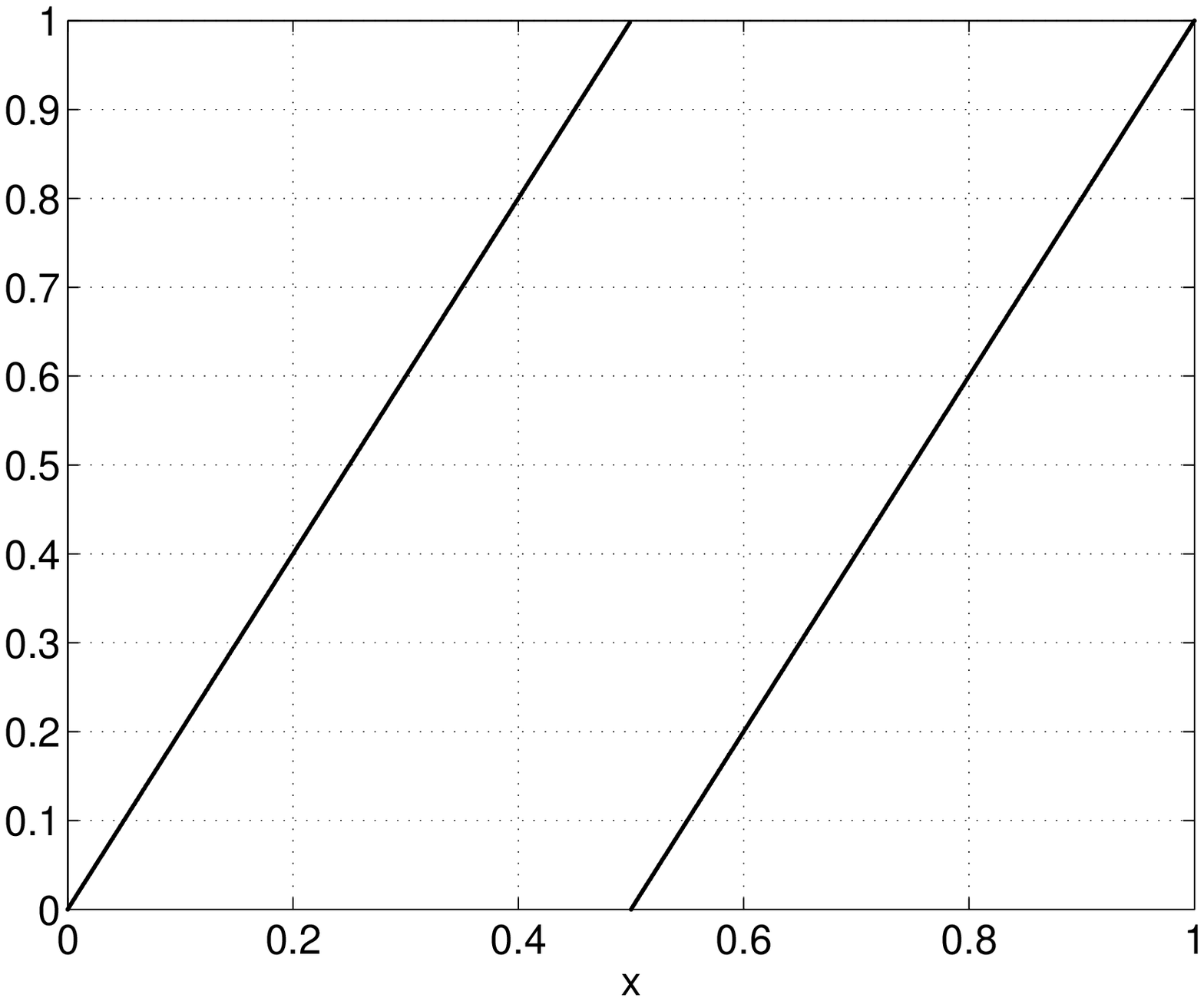}
\includegraphics[scale=0.25]{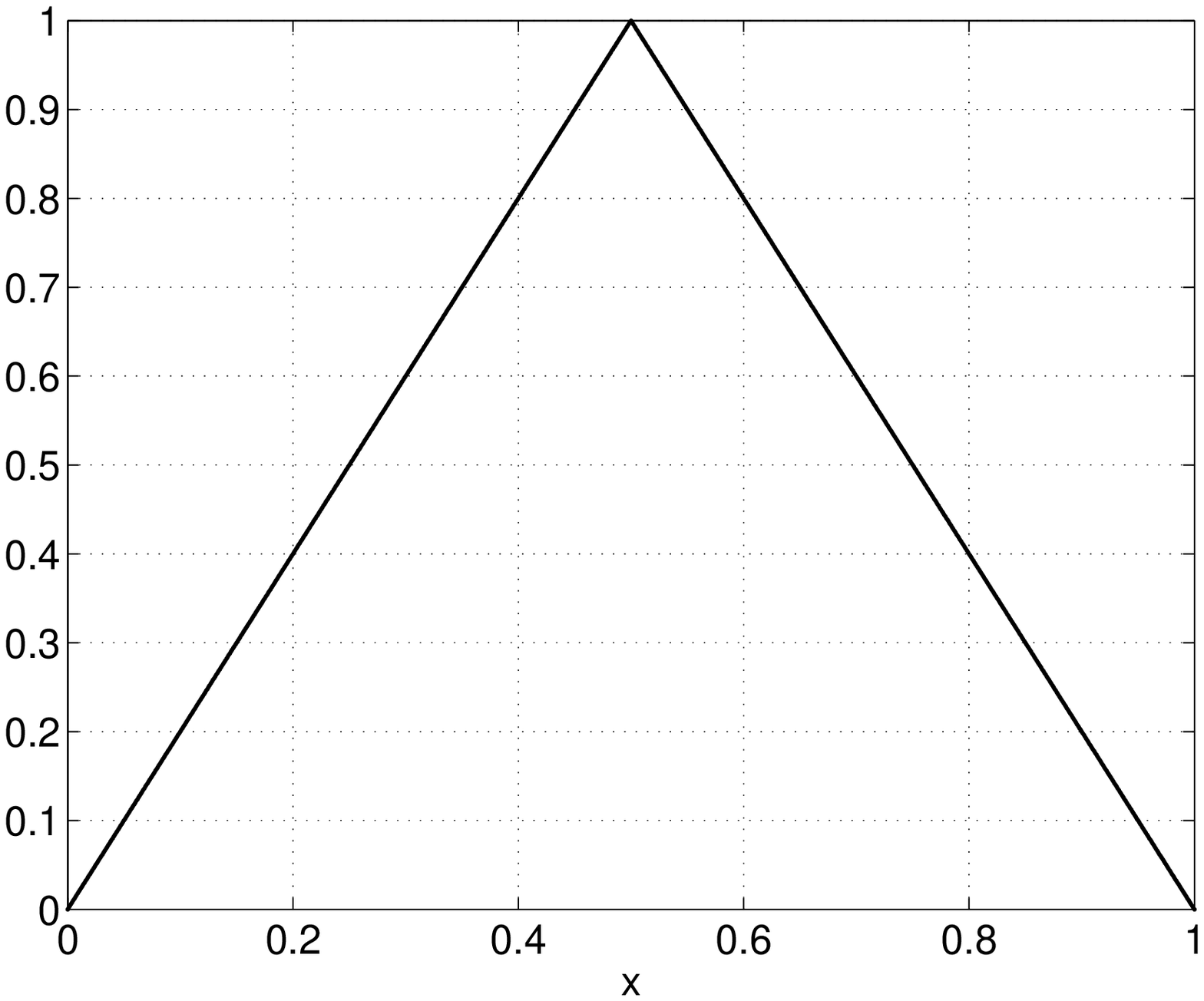}\\
(a)~~~~~~~~~~~~~~~~~~~~~~~~~~~~~~~~~~ (b)
 \caption[Some well known GLS]{Some well known GLS: (a)
Standard Binary map ($x \mapsto 2x$, $0 \leq x < 0.5$; $x \mapsto
2x-1$, $0.5 \leq x < 1$). (b) Standard Tent map ($x \mapsto 2x$, $0
\leq x < 0.5$; $x \mapsto 2-2x$, $0.5 \leq x < 1$).}
\label{figure:wellknownGLS}
\end{figure}

It is property 2 and 3 that allow a faithful ``embedding'' of a
stochastic i.i.d source. For a proof of these properties, please
refer Dajani et. al.~\cite{Dajani2002}. Some well known GLS are the
standard Binary map and the standard Tent map shown in
Figure~\ref{figure:wellknownGLS}.

%\begin{figure}[!h]
%\centering
%\begin{minipage}[t]{\figwidth}
%\centering
%\includegraphics[width=\textwidth]{binary1}
%a) Standard Binary map ($x \mapsto 2x$, $0 \leq x < 0.5$; $x \mapsto
%2x-1$, $0.5 \leq x < 1$).
%\end{minipage}
%\begin{minipage}[t]{\figwidth}
%\centering
%\includegraphics[width=\textwidth]{tent1}
%b) Standard Tent map \\($x \mapsto 2x$, $0 \leq x < 0.5$; $x \mapsto
%2-2x$, $0.5 \leq x < 1$).
%\end{minipage}
%\caption[Some well known GLS]{Some well known GLS.}
%\label{figure:wellknownGLS}
%\end{figure}

\subsection{Lyapunov Exponent of GLS = Shannon's Entropy}
It is easy to verify that GLS preserves the Lebesgue measure. A
probability density $\Pi(x)$ on [0,1) is invariant under the given
transformation $T$, if for each interval $[c,d] \subset [0,1)$, we
have:

\begin{equation}
\int_c^d \Pi(x)dx = \int_{\mathcal{S}} \Pi(x)dx.
\end{equation}
where $\mathcal{S} = T^{-1}([c,d]) = \{ x | c \leq T(x) \leq d \}$.

For the GLS, the above condition has constant probability density on
$[0,1)$ as the only solution. It then follows from Birkhoff's
ergodic theorem~\cite{Dajani2002} that the asymptotic probability
distribution of the points of almost every trajectory is uniform. We
can hence calculate Lyapunov exponent as follows:

\begin{equation}
\lambda = \int_0^1 log_2(|T'(x)|) \Pi(x)dx.~~~\textrm{(almost
everywhere)}
\end{equation}
Here, we measure $\lambda$ in bits/iteration.

\begin{figure}[!h]
\centering
\includegraphics[scale=.3]{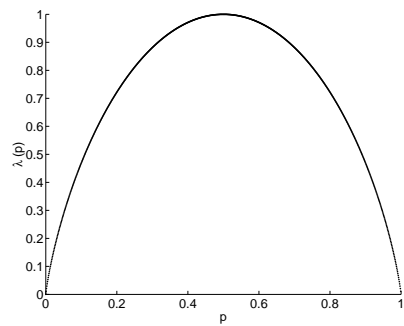}
\caption[Plot of Lyapunov exponent for binary i.i.d
sources]{Lyapunov Exponent: $\lambda (p) = -p\log_2(p)
-(1-p)\log_2(1-p)$ in units of bits/iteration plotted against $p$
for binary i.i.d sources. The maximum occurs at $p=\frac{1}{2}$.
Note that $\lambda (p) = \lambda (1-p)$. } \label{fig:figHp}
\end{figure}

$\Pi(x)$ is uniform with value 1 on [0,1) and $T'(x) = constant$
since $T(x)$ is linear in each of the intervals, the above
expression simplifies to:

\begin{equation}
\lambda = -\sum_{i=1,p_i\neq0}^{i=N} p_ilog_2(p_i).~~~
\textrm{(almost everywhere)}
\end{equation}
This turns out to be equal to Shannon's entropy of the i.i.d source
$X$. Thus Lyapunov exponent of the GLS that faithfully embeds the
stochastic i.i.d source $X$ is equal to the Shannon's entropy of the
source. Lyapunov exponent can be understood as the amount of
information in bits revealed by the symbolic sequence (measurement)
of the dynamical system in every iteration~\footnote{This is equal
to the Kolmogorov-Sinai entropy which is defined as the sum of
positive Lyapunov exponents. In a 1D chaotic map, there is only one
Lyapunov exponent and it is positive.}. It can be seen that the
Lyapunov exponent for all the modes of the GLS are the same. The
Lyapunov exponent for binary i.i.d sources is plotted in
Figure~\ref{fig:figHp} as a function of $p$ (the probability of
symbol `0').

\section{Successive Source Approximation using GLS}
In this section, we address the measurement problem proposed in
Section~\ref{Section:Measurement Problem}. Throughout our analysis,
$N > 2$ (finite) and $M=2$ is assumed. We are seeking {\it
minimum-redundancy
binary symbol} codes. ``Minimum-redundancy'' is defined as follows~\cite{Huffman1952}:\\

\begin{defn2}[Minimum Redundancy]
A binary symbol code $\mathcal{C} = \{ c_1, c_2, \ldots, c_N \}$
with lengths $L = \{ l_1, l_2, \ldots, l_N \}$ for the i.i.d source
$X$ with alphabet $\mathcal{A} = \{a_1, a_2, \ldots, a_N \}$ with
respective probabilities $\{p_1, p_2, \ldots, p_N \}$ is said to
have minimum-redundancy if $L_{\mathcal{C}}(X) = \sum_{i=1}^{N} l_i
p_i$ is minimum.
\end{defn2}

For $N=2$, the minimum-redundancy binary symbol code for the
alphabet $\mathcal{A} = \{a_1, a_2\}$ is $\mathcal{C} = \{ 0, 1 \}$
($a_1 \mapsto 0$, $a_2 \mapsto 1$). The goal of source coding is to
minimize $L_{\mathcal{C}}(X)$, the average code-word length of
$\mathcal{C}$, since this is important in any communication system.
As we mentioned before, it is always true that $L_{\mathcal{C}}(X)
\geq H(X)$~\cite{Shannon1948}.

\subsection{Successive Source Approximation Algorithm using GLS}
Our approach is to approximate the original i.i.d source (GLS with
$N$ partitions) with the {\it best} GLS with a reduced number of
partitions (reduced by 1). For the sake of notational convenience,
we shall term the original GLS as order $N$ (for original source
$X_N$) and the reduced GLS would be of order $N-1$ (for
approximating source $X_{N-1}$). This new source $X_{N-1}$ is now
approximated further with the {\it best} possible source of order
$N-2$ ($X_{N-2}$). This procedure of successive approximation of
sources is repeated until we end up with a GLS of order $M = 2$
($X_2$). It has only two partitions for which we know the
minimum-redundancy symbol code is $\mathcal{C} = \{ 0, 1 \}$.

At any given stage $q$ of approximation, the easiest way to
construct a source of order $q-1$ is to merge two of the existing
$q$ partitions. What should be the rationale for determining which
is the {\it best} $q-1$ order approximating source $X_{q-1}$ for the
source $X_q$?

\begin{defn2}[Best Approximating Source]
%\caption{Best Approximating Source} \label{defn:BestApprox} %
Among all possible $q-1$ order approximating sources, the best
approximation is the one which minimizes the following quantity:
\begin{equation}
\Delta = \lambda(X_{q}) - \lambda(X_{q-1}).
\end{equation}
\end{defn2}

where $\lambda(\cdot)$ is the Lyapunov exponent of the argument. The
reason behind this choice is intuitive. We have already established
that the Lyapunov exponent is equal to the Shannon's entropy for the
GLS and that it represents the amount of information (in bits)
revealed by the symbolic sequence of the source at every iteration.
Thus, the best approximating source should be as close as possible
to the original source in terms of Lyapunov exponent.

There are three steps to our algorithm for finding minimum
redundancy binary symbol code as given below here:

\begin{algorithm}[!h]
\caption{Successive Source Approximation using GLS} \label{alg:SuccSourceGLS} %
\begin{enumerate}
\item Embed the i.i.d source $X$ in to a GLS with $N$ partitions as described in~\ref{Subsection:embedding}. Initialize $K=N$. The source is denoted by $X_K$ to indicate
order $K$.

\item Approximate source $X_K$ with a GLS with $K-1$ partitions by merging the {\it smallest} two partitions to
obtain the source $X_{K-1}$ of order $K-1$. $K \leftarrow K-1$.

\item Repeat step 2 until order of the GLS is 2 ($K=2$), then, stop.
\end{enumerate}
\end{algorithm}

We shall prove that the approximating source which merges the two
{\it smallest} partitions is the {\it best} approximating source. It
shall be subsequently proved that this algorithm leads to
minimum-redundancy, i.e., it minimizes $L_{\mathcal{C}}(X)$.
Assigning codewords to the alphabets will also be shown.\\

\noindent {\bf Theorem 1: (Best Successive Source Approximation)}
{\it For a source $X_M$ which takes values from $\{A_1, A_2, \ldots
A_{M-1}, A_M \}$ with probabilities $\{ p_1, p_2, \ldots p_{M-1},
p_M \}$ respectively and with $1 \geq p_1 \geq p_2 \geq \ldots \geq
p_{M-1} \geq p_M \geq 0$ ($\sum_{i=1}^{i=M} p_i = 1$), the source
$X_{M-1}$ which is the {\bf best} $M$-1 order approximation to $X_M$
has probabilities $\{ p_1, p_2, \ldots p_{M-2}, p_{M-1}+p_M \}$.}\\

\noindent{\bf Proof:\\} By induction on $M$. For $M=1$ and $M=2$,
there is nothing to prove. We will first show that the statement is
true for $M=3$.

\begin{figure} [!h]
\centering
\includegraphics[scale=0.5]{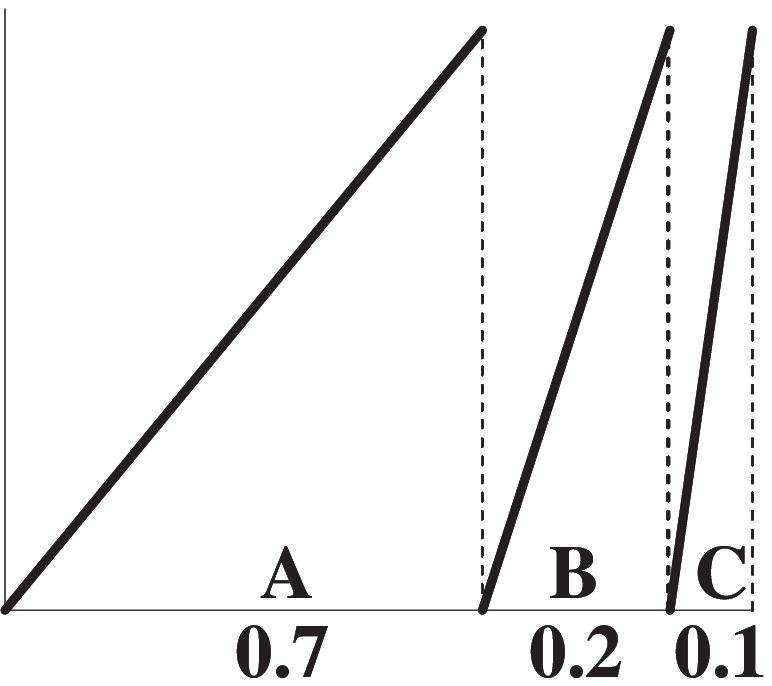}\\
(a) Source $X$: \{$A$,$B$,$C$\} with probabilities \{0.7, 0.2, 0.1\}, $\lambda_{X} = 1.156$. \\
\vspace{0.1in}
\includegraphics[scale=0.5]{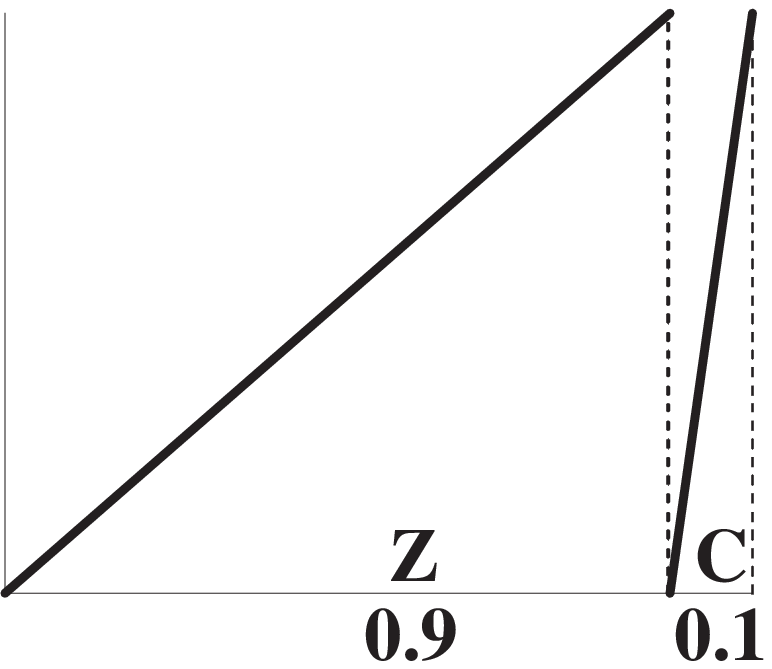} %[width=\textwidth]
\hspace{0.1in}
\includegraphics[scale=0.5]{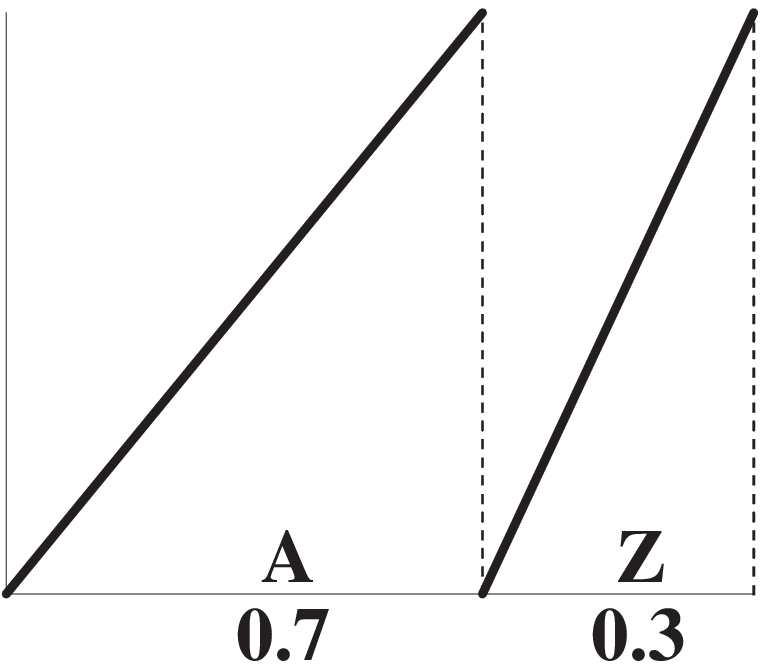}
\hspace{0.1in}\\

(b) $A$ and $B$ merged. ~~~~~~~~~~~~~~ (c) $B$ and $C$ merged.\\
($\lambda_{AB} = 0.47$) ~~~~~~~~~~~~~~~~~~~~~~~~~~~~($\lambda_{BC} =
0.88$)

\includegraphics[scale=0.5]{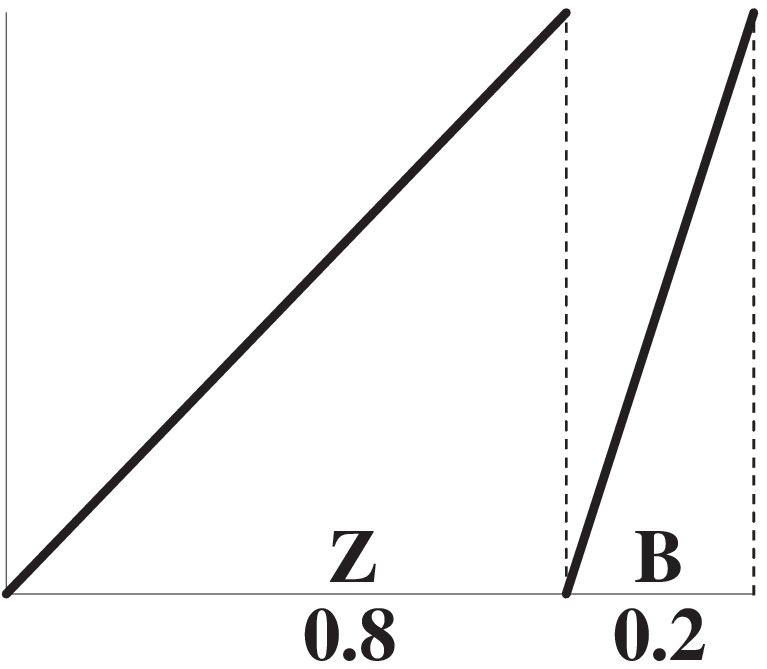}\\
(d) $A$ and $C$ merged ($\lambda_{AC} = 0.73$).
%~~~~~~~b) $A$ and $B$ merged.~~~~~~~ c) $B$ and $C$ merged.~~~~~~~ d) $A$ and $C$ merged.~~~~~~~ \\%
%($\lambda_{AB} = 0.47$)~~~~~~~~~~~~~~  ($\lambda_{BC} = 0.88$)~~~~~~~~~~~~~~~~~~  ($\lambda_{AC} = 0.73$)%
\caption[Successive source approximation using GLS: An
example]{Successive source approximation using GLS: An example.
Here, $\lambda_{BC}$ is the closest to $\lambda_{X}$. Unit of
$\lambda(.)$ is bits/iteration. } \label{figure:Huffman1}
\end{figure}

\begin{itemize}
\item {\bf $M=3$}. $X_3$ takes values from $\{a_1, a_2, a_3 \}$ with
probabilities $\{p_1, p_2, p_3 \}$ respectively and $1 \geq p_1 \geq
p_2 \geq p_3 \geq 0$ with $p_1 + p_2 + p_3 = 1$.\\

We need to show that $X_2$ which takes values from $\{a_1, Z\}$ with
probabilities $\{p_1, p_2+p_3 \}$ is the best $2$-order
approximation to $X_3$. Here $Z$ is a symbol that represents the merged partition.\\

This means, that we should show that this is better than any other
$2$-order approximation. There are two other $2$-order
approximations, namely, $Y_2$ which takes values from $\{ a_3,Z \}$
with probabilities $\{ p_3, p_2+ p_1 \}$ and $W_2$ which takes
values from $\{a_2, Z \}$ with probabilities $\{ p_2,p_1+p_3 \}$. \\

This implies that we need to show $\lambda(X_3)-\lambda(X_2) \leq
\lambda(X_3)-\lambda(Y_2)$ and $\lambda(X_3)-\lambda(X_2) \leq
\lambda(X_3)-\lambda(W_2)$.

\item We shall prove $\lambda(X_3)-\lambda(X_2) \leq
\lambda(X_3)-\lambda(Y_2)$.

This means that we need to prove $\lambda(X_2) \geq \lambda(Y_2)$.
This means we need to show $-p_1 log_2(p_1) -(p_2+p_3)
log_2(p_2+p_3) \geq -p_3 log_2(p_3) -(p_1+p_2) log_2(p_1+p_2)$. We
need to show the following:
\begin{eqnarray*}
-p_1 log_2(p_1) - (1-p_1) log_2(1-p_1) &\geq& -p_3 log_2(p_3)\\
& &  -(1-p_3) log_2(1-p_3)\\
\Rightarrow \lambda_2(p_1) &\geq& \lambda_2(p_3).
\end{eqnarray*}
There are two cases. If $p_1 \leq 0.5$, then since $p_3 \leq p_1$,
$\lambda_2(p_1) \geq \lambda_2(p_3)$. If $p_1 > 0.5$, then since
$p_2+p_3 = 1-p_1$, we have $p_3 \leq 1-p_1 $. This again implies
$\lambda_2(p_1) \geq \lambda_2(p_3)$. Thus, we have proved that
$X_2$ is better than $Y_2$.

\item We can follow the same argument to prove that $\lambda(X_2) \geq
\lambda(W_2)$. Thus, we have shown that the theorem is true for
$M=3$. An illustrated example is given in
Figure~\ref{figure:Huffman1}.

\item Induction hypothesis: Assume that the theorem is true for
$M=k$, we need to prove that this implies that the theorem is true
for $M=k+1$.\\

Let $X_{k+1}$ have the probability distribution $\{ p_1, p_2, \ldots
p_k,p_{k+1}\}$. Let us assume that $p_1 \neq 1$ (if this is the
case, there is nothing to prove). This means that $1-p_1 > 0$.
Divide all the probabilities by ($1-p_1$) to get
$\{\frac{p_1}{1-p_1}, \frac{p_2}{1-p_1}, \frac{p_3}{1-p_1} \ldots
\frac{p_k}{1-p_1}, \frac{p_{k+1}}{1-p_1} \}$. Consider the set $\{
\frac{p_2}{1-p_1}, \frac{p_3}{1-p_1} \ldots \frac{p_k}{1-p_1}
,\frac{p_{k+1}}{1-p_1} \}$. This represents a probability
distribution of a source with $k$
possible values and we know that the theorem is true for $M=k$.\\

This means that the best source approximation for this new
distribution is a source with probability distribution $\{
\frac{p_2}{1-p_1}, \frac{p_3}{1-p_1} \ldots
\frac{p_k+p_{k+1}}{1-p_1} \}$.\\

In other words, this means:
\begin{equation*}
-\sum_{i=2}^{k-1}(\frac{p_i}{1-p_1})log_2(\frac{p_i}{1-p_1})
-(\frac{p_k+p_{k+1}}{1-p_1})log_2(\frac{p_k+p_{k+1}}{1-p_1})\\
\end{equation*}
\begin{eqnarray*} &\geq&
\end{eqnarray*}
\begin{equation*}
-\sum_{i=2,i \neq r, i\neq
s}^{k+1}(\frac{p_i}{1-p_1})log_2(\frac{p_i}{1-p_1})
-(\frac{p_r+p_s}{1-p_1})log_2(\frac{p_r+p_s}{1-p_1}).
\end{equation*}

where $r$ and $s$ are both different from $k$ and $k+1$. Multiply on
both sides by $(1-p_1)$ and simplify to get:
\begin{eqnarray*}
-\sum_{i=2}^{k-1} p_i log_2(p_i) -(p_k+p_{k+1})log_2(p_k+p_{k+1})
&\geq&
\\
-\sum_{i=2,i \neq r, i\neq s}^{k+1} p_i log_2(p_i)
-(p_r+p_s)log_2(p_r+p_s).
\end{eqnarray*}
Add the term $-p_1log_2(p_1) > 0$  on both sides and we have proved
that the {\bf best} $k$-order approximation to $X_{k+1}$ is the
source $X_k$, where symbols with the two least probabilities are
merged together. We have thus proved the theorem.$\hfill \square$
\end{itemize}

\subsection{Codewords are Symbolic Sequences}
%Having noted that our algorithm is a re-discovery of the popular
%Huffman coding algorithm, we are in a position to describe the code
%$\mathcal{C}$. It has to essentially agree with the Huffman code for
%Theorem~\ref{Thm:SSA1} to hold.

At the end of Algorithm~\ref{alg:SuccSourceGLS}, we have order-2
approximation ($X_2$). We allocate the code $C_2 = \{0,1\}$ to the
two partitions. When we go from $X_2$ to $X_3$, the two sibling
partitions that were merged to form the parent partition will get
the codes `$S0$' and `$S1$' where `$S$' is the codeword of the
parent partition. This process is repeated until we have allocated
codewords to $X_N$.

%We then have the code $C_N$ for $X_N$ which agrees with the Huffman
%code.

It is interesting to realize that the codewords are actually
symbolic sequences on the standard binary map. By allocating the
code $C_2 = \{0,1\}$ to $X_2$ we are essentially treating the two
partitions to have equal probabilities although they may be highly
skewed. In fact, we are approximating the source $X_2$ as a GLS with
equal partitions (=0.5 each) which is the standard binary map. The
code $C_2$ is thus the symbolic sequence on the standard binary map.
Now, moving up from $X_2$ to $X_3$ we are doing the same
approximation. We are treating the two sibling partitions to have
equal probabilities and giving them the codes `$S0$' and `$S1$'
which are the symbolic sequences for those two partitions on the
standard binary map. Continuing in this fashion, we see that all the
codes are symbolic sequences on the standard binary map. Every
alphabet of the source $X$ is {\it approximated} to a partition on
the binary map and the codeword allocated to it is the corresponding
symbolic sequence. It will be proved that the approximation is
minimum redundancy and as a consequence of this, if the
probabilities are all powers of 2, then the approximation is not
only minimum redundancy but also equals the entropy of the source
($L_{\mathcal{C}}(X) = H(X)$).\\

\noindent {\bf Theorem 2: (Successive Source Approximation)} {\it
The successive source approximation algorithm using GLS yields
minimum-redundancy (i.e., it minimizes $L_{\mathcal{C}}(X)$).}\\

\noindent{\bf Proof:\\} We make the important observation that the
successive source approximation algorithm is in fact a re-discovery
of the binary Huffman coding algorithm~\cite{Huffman1952} which is
known to minimize $L_{\mathcal{C}}(X)$ and hence yields
minimum-redundancy. Since our algorithm is essentially a
re-discovery of the binary Huffman coding algorithm, the theorem is
proved (the codewords allocated in the previous section are the same
as Huffman codes). $\hfill \square$

\subsection{Encoding and Decoding}
We have described how by successively approximating the original
stochastic i.i.d source using GLS, we arrive at a set of codewords
for the alphabet which achieves minimum redundancy. The assignment
of symbolic sequences as codewords to the alphabet of the source is
the process of encoding. Thus, given a series of observations of
$X$, the measuring device represents and stores these as codewords.
For decoding, the reverse process needs to be applied, i.e., the
codewords have to be replaced by the observations. This can be
performed by another device which has a look-up table consisting of
the alphabet set and the corresponding codewords which were assigned
originally by the measuring device.

%
%%Although Huffman codes are the {\it best (minimum-redundancy)}
%%symbol codes, but to achieve true optimality ($H(X)$), the block
%%size has to be very large. As the block size $B$ increases, the
%%number of alphabets to handle exponentially increases ($2^B$). This
%%poses serious complexity concerns and is a major limitation of the
%%scheme. Furthermore, for decoding to begin, the Huffman code has to
%%be available at the decoder. The overhead bits necessary for this
%%grows with increasing block-size as is evident in the experimental
%%results provided. This compensates for the gain in compression ratio
%%by blocking. Thus, effectively we lose out both on compression ratio
%%and complexity.

\section{Some Remarks}
We make some important observations/remarks here:

\begin{enumerate}
\item The faithful modeling of a stochastic i.i.d source as a GLS is
a very important step. This ensured that the Lyapunov exponent
captured the information content (Shannon's Entropy) of the source.

\item Codewords are symbolic sequences on GLS. We could have chosen
a different scheme for giving codewords than the one described here.
For example, we could have chosen symbolic sequences on the Tent map
as codewords. This would also correspond to a different set of
Huffman codes, but with the same average codeword length
$L_{\mathcal{C}}(X)$. Huffman codes are not unique but depend on the
way we assign codewords at every level.

\item Huffman codes are {\it symbol codes}, i.e., each symbol in the
alphabet is given a distinct codeword. We have investigated binary
codes in this paper. An extension to the proposed algorithm is
possible for ternary and higher bases.

\item In another related work, we have used GLS to design {\it stream codes}.
Unlike symbol codes, stream codes encode multiple symbols at a time.
Therefore, individual symbols in the alphabet no longer correspond
to distinct codewords. By treating the entire message as a symbolic
sequence on the GLS, we encode the initial condition which contains
the same information. This achieves optimal lossless compression as
demonstrated in~\cite{NithinGLS}.

\item We have extended GLS to piecewise non-linear, yet Lebesgue
measure preserving discrete chaotic dynamical systems. These have
very interesting properties (such as Robust Chaos in two parameters)
and are useful for joint compression and encryption
applications~\cite{NithinGLS}.

\end{enumerate}

\section{Conclusions}
Source coding problem is motivated as a measurement problem. A
stochastic i.i.d source can be faithfully ``embedded'' into a
piecewise linear chaotic dynamical system (GLS) which exhibits
interesting properties. The Lyapunov exponent of the GLS is equal to
Shannon's entropy of the i.i.d source. The measurement problem is
addressed by successive source approximation using GLS with the
nearest Lyapunov exponent (by merging the two least probable
states). By assigning symbolic sequences as codewords, we
re-discovered the popular Huffman coding algorithm -- a minimum
redundancy symbol code for i.i.d sources.

\section*{Acknowledgements}
Nithin Nagaraj is grateful to Prabhakar G. Vaidya and Kishor G. Bhat
for discussions on GLS. He is also thankful to the Department of
Science and Technology for funding this work as a part of the Ph.D.
program at National Institute of Advanced Studies, Indian Institute
of Science Campus, Bangalore. The author is indebted to Nikita
Sidorov, Mathematics Dept., Univ. of Manchester, for providing
references to Cantor's work.

%\appendix
%\section{Appendixes}
%\newpage %Just because of unusual number of tables stacked at end

%%%%\bibliography{ac_chaos}% Produces the bibliography via BibTeX.
%%%%%%%%%%%%%%%%%%%%%%%%%%%%%%%%%%%%%%%%%%%%%%%%%%%%%%%%%%%%%%%%
%%%%%%%%%%%%%%%%%%%%%%%%%%%%%%%%%%%%%%%%%%%%%%%%%%%%%%%%%%%%%%%%

%%%%%%%%%%%%%%%%%%%%%%%%%%%%%%%%%%%%%%%%%%%%%%%%%%%%%%%%%%%%%%%%

\end{document}